\documentclass[conference]{IEEEtran}
\IEEEoverridecommandlockouts
\usepackage{cite}
\usepackage{amsmath,amssymb,amsfonts}
\usepackage{algorithmic}
\usepackage{listings}
\usepackage{graphicx}
\usepackage{textcomp}
\usepackage{url}
\usepackage{lipsum}
\usepackage[nottoc, notlof, notlot]{tocbibind}

\usepackage{color}
\usepackage[type={CC},modifier={by-nc-sa},version={4.0},]{doclicense} % Tu peux jouer sur la licence CC que tu veux.

\DeclareFixedFont{\ttb}{T1}{txtt}{bx}{n}{12} % for bold
\DeclareFixedFont{\ttm}{T1}{txtt}{m}{n}{12}  % for normal
\definecolor{deepblue}{rgb}{0,0,0.5}
\definecolor{deepred}{rgb}{0.6,0,0}
\definecolor{deepgreen}{rgb}{0,0.5,0}

\begin{document}
	\lstset{
		language=Python,
		basicstyle=\footnotesize ,%\tiny, \small \footnotesize etc.
		%basicstyle=\ttm,
		otherkeywords={self},             % Add keywords here
		keywordstyle=\color{deepblue},
		keepspaces=true,
		emph={MyClass,__init__},          % Custom highlighting
		%emphstyle=\ttb\color{deepred},    % Custom highlighting style
		%stringstyle=\color{deepgreen},
		frame=tb,                         % Any extra options here
		showstringspaces=false            %
	}

	\title{LiteX: an open-source SoC builder and library based on Migen Python DSL
	\thanks{\doclicenseText\doclicenseImage[imagewidth=6em]}}

	\author{
		\IEEEauthorblockN{Florent Kermarrec}
		\IEEEauthorblockA{\textit{Enjoy-Digital, Landivisiau, France} \\
			florent@enjoy-digital.fr}

		\and
		\IEEEauthorblockN{S\'ebastien Bourdeauducq}
		\IEEEauthorblockA{\textit{M-Labs, Hong-Kong} \\
			sb@m-labs.hk}

		\and
		\IEEEauthorblockN{Jean-Christophe Le Lann and Hannah Badier}
		\IEEEauthorblockA{\textit{ENSTA Bretagne, Brest and Lab-STICC UMR } \\
			lelannje@ensta-bretagne.fr}
	}

	\maketitle

	\begin{abstract}
		LiteX\cite{LiteX} is a GitHub-hosted SoC builder / IP library and utilities that can be used to create SoCs and full FPGA designs.
		Besides being open-source and BSD licensed, its originality lies in the fact that its IP components are entirely described using
		Migen Python internal DSL, which simplifies its design in depth.
		LiteX already supports various softcores CPUs and essential peripherals, with no dependencies on proprietary
		IP blocks or generators.
		This paper provides an overview of LiteX: two real SoC designs on FPGA are presented.
		They both leverage the LiteX approach in terms of design entry, libraries and integration capabilities.
		The first one is based on RISC-V core, while the second is based on a LM32 core.
		In the second use case, we further demonstrate the use of a fully open-source toolchain coupled with LiteX.\end{abstract}
	\begin{IEEEkeywords}
		open-source, SoC, Python, DSL, FPGA, IP library
	\end{IEEEkeywords}

	\section{Introduction}
	% EDA is important.
	%\subsection{EDA \& HDLs: an quick historical perspective}
	Electronic Design Automation (EDA) plays a central role in the advent of all the electronic devices we use daily, ranging from small embedded systems to internet-scale infrastructures.
	The success of EDA may mainly be attributed to its capability to describe semi-conductor devices with tools and languages based on robust abstraction layers:
	transistors, layout, logic, register-transfer level (RTL), behavioral descriptions, etc.
	Thanks to these levels of abstraction, engineers do not have to face the entire complexity of the underlying device they are designing, which makes the design process more constructive and efficient.
	In this process, Hardware Description Languages (HDLs), such as Verilog and VHDL, still play a pivotal role. New features that aim at facilitating the designer's job have been added at a slow pace. This conservative slowness has left only very little room for newcomers. Among them, SystemVerilog has succeeded to emerge as the most natural way to describe and execute complex test benches, while SystemC is now mainly associated to transaction-level platform modeling.
	On the contrary, synchronous languages \cite{benveniste2003synchronous}, whose formal aspects could have been beneficial to EDA, have not been adopted widely in EDA, despite decades of extensive research.
	In the same vein, High-level Synthesis, which calls for C-based design entry, has not gained the popularity of HDLs yet.
	Nowadays, classical HDLs, acting as RTL description languages, still stand as the cornerstone of EDA.
	% FPGA dope l'acvitite et ouvre des voies d'innovation.

	% DSL
	%\subsection{Internal DSLs principles}
	\subsection{EDA renewal through DSLs and platform-based design}
	A recent trend, however, shows that several new HDLs are starting to emerge. Their common characteristic is that they are essentially internal Domain Specific Languages (DSLs). These languages --sometimes called ``embedded DSLs''--
	are hosted by a mainstream programming language. No matter which host language is used, embedding a DSL in a mainstream programming language provides several benefits:
	comfort, smooth learning curve, access to host language libraries and ecosystem. Complex compiler technologies or specific parsers are not necessary, which significantly reduces the amount of work required compared to the effort needed to create a brand-new language.
	The design of such internal DSLs is based on efficiently mapping domain-specific concepts
	to the host language syntax.
	%A good naming of functions in a procedural language (like C) may be seen as starting point %when designing
	%an embryonic DSL : in the software engineering folklore, Programming amounts to create %one own's DSL" (this citation if attributed to Dave Thomas).
	%Some mainstream languages are known to be particularly well suited for this hosting, due %to the malleability of their syntax,
	%up to the point where it may be hard to distinguish between the host and the hosted %languages.
	%This is the case of modern object-oriented languages such as Ruby, known for web-oriented %frameworks like Rails, or Groovy, that permits a
	%simplified access to their abstract syntax tree (AST) programmatically or %meta-programmatically. But Lisp --which is a functional language--
	%may be seen as the main influencer of this trend : the core syntax of Lisp is {\it %s-expressions} that directly represent the AST.
	%Programming a simple function in Lisp amounts to create a new s-expression, which gives %the illusion of augmenting the language itself.
	%The ideal host language should offer, beyond its own advantages, such ease of description.

	%However, as defended in this paper, the success of an embedded DSL rests on several %other aspects, particularly on its libraries.

	The renewal of HDLs, based on embedded DSLs like Chisel/SpinalHDL, is also encouraged by the advent of RISC-V platform-based design (PBD).
	This initiative is intended to counterbalance the supremacy of ARM in the field of embedded systems, by providing a set of royalty-free softcores. Such softcores play a central role in platform-based design, which has been defined \cite{Bailey2010} as ``an integration oriented design approach emphasizing
	systematic reuse, for developing complex products based upon platforms and compatible hardware and software virtual components,
	intended to reduce development risks, costs and time to market''. This reuse calls for new methodologies: to build complex
	embedded systems organized around a processor, two ingredients are mandatory. The first is abstraction: modern languages are likely to provide such characteristics though either object-oriented or
	functional modeling. As clearly stated in the previous definition, the second ingredient is the availability of component libraries.

	\subsection{Objectives of the paper}
	In this paper, we present LiteX\cite{LiteX}, SoC builder and library of IP components described at the RTL level, together with various utilities that facilitate the
	building of complete SoC designs. LiteX resorts to a DSL, written in Python and
	named Migen. LiteX has been used successfully and on a daily basis for several years by Enjoy-Digital \cite{EnjoyDigital}, a company dedicated to open-source FPGA-based design,  in various application domains
	(multimedia, software defined radio, automotive, etc.) and is progressively being adopted by users around the world.
	The obvious key of this adoption lies both in the very permissive open-source license adopted by LiteX (BSD), and the choice of Python for the RTL descriptions and deployment.
	LiteX participates in a larger trend that will be discussed in this paper.

	The rest of the paper is organized as follows. Next Section makes a short survey of the trend of eDSLs for EDA, and especially for RTL design.
	Section 3 presents Migen FHDL and MiSoC. Section 4 introduces LiteX SoC builder and library. In Section 5, a full SoC design on FPGA is exposed and a fully open-source design flow based on LiteX is presented.

	\section{Related work : benefits of internal DSLs for RTL Design}
	Many hardware-oriented embedded DSLs have been proposed.
	%Table tries to summerize these proposals.
	Among them, SystemC itself can be cited at first. It was proposed by Liao in \cite{Liao97} (at that time, SystemC was named Scenic).
	SystemC relies on C++ objects and allows to describe hardware systems at various level of abstractions.
	The simulation engine remains event-driven, similar to VHDL and Verilog.
	It has received a large audience, and has improved since the early paper of Liao to include sophisticated features for transaction-level modeling \cite{bombieri09},
	which allows to build models usable by
	software programmers for early SoC validation. However, SystemC has not been used fully as RTL entry.

	A DSL that has recently been receiving growing attention is Chisel \cite{bachrach12} and its fork named SpinalHDL.
	They both strongly contribute to the recent renew of RTL design practices. They also take an active part in the advent of RISC-V softcores.
	Chisel initially targets synchronous designs, where the clock is implicit. The host language is Scala, a multi-paradigm language
	(object-oriented and functional), with a static typing.
	Here again, Scala \cite{OderskyR14} is known for its syntactic flexibility: semicolons are optional, any method can be used as an infix operator, etc.
	These features provide a strong differentiator with respect to VHDL for instance, which is known to be extremely verbose.
	However, the benefits of Scala go far beyond the syntax: the type inference mechanism, parameterized types and generators are particularly appealing for
	hardware design. HardCaml, an OCaml library for designing hardware, shares several aspects with Chisel, as OCaml offers many of the same modern multi-paradigm features provided by Scala.
	Haskell (a pure functional language) has also been used extensively for describing hardware:
	Lava \cite{bjesse1998lava},
	$C\lambda aSH$ \cite{baaij10}, \cite{Zhai15}. An older approach based on Java for FPGA programming was proposed in\cite{guccione1999jbits}. Furthermore, an interesting approach named MyHDL\cite{JaicS15} and based on Python was introduced by Decaluwe in \cite{myhdl}. It leverages the interpreted nature of Python and shares several characteristics with Migen.

	\section{Migen}

	\begin{figure*}[htb]
		\centering
		\includegraphics[scale=0.3]{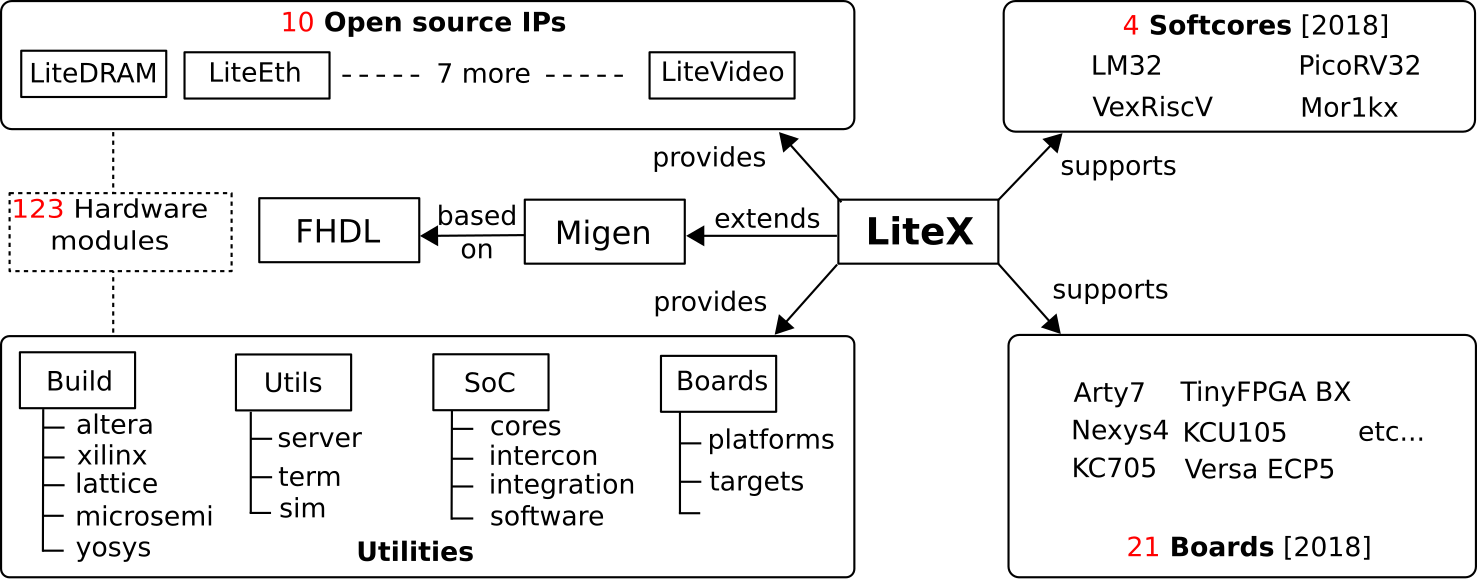}
		\caption{An overview of LiteX framework}
		\label{litex}
	\end{figure*}

	Migen, a fully open-source Python-based toolbox, contains a HDL, a library of cores, a simulator and a build system. Migen originated from Milkymist project and stands for ``Milkymist generator''.

	\subsection{Migen FHDL}
	Similar to previous cited technologies, Migen FHDL (Fragmented Hardware Description Language) allows to describe and simulate RTL circuits. It is based on a custom Python abstract syntax tree (AST) and can produce synthesizable Verilog. FHDL does not follow the event-driven paradigm of most HDLs, and instead replaces it with notions of synchronous and combinatorial statements.

	Thanks to Migen FHDL, highly and easily configurable cores can be designed. Creating a design by writing a Python program raises the level of abstraction, by for example enabling the use of object-oriented programming or meta-programming. The choice of Python in particular, as opposed to other programming languages such as Scala, also gives a clear advantage for Migen adoption: Python is an easy to learn and well-known language, already used for many other tasks by application engineers, for instance for algorithmic prototyping.

	The following code example shows how a functional design can be implemented in a short, efficient and readable manner:

	\begin{lstlisting}[frame=single,caption={Blinking LED design example},captionpos=b]
from migen import *
from migen.fhdl import verilog

class Blinker(Module):
  def __init__(self, sys_clk_freq, period):
    self.led = led = Signal()

    # # #

    toggle = Signal()
    counter_preload = int(sys_clk_freq*period/2)
    counter = Signal(max=counter_preload + 1)

    self.comb += toggle.eq(counter == 0)
    self.sync += \
    If(toggle,
      led.eq(~led),
      counter.eq(counter_preload)
    ).Else(
      counter.eq(counter - 1)
    )

# Create a 10Hz blinker from a 100MHz system clock.
blinker = Blinker(sys_clk_freq=100e6, period=1e-1)
print(verilog.convert(blinker, {blinker.led}))

	\end{lstlisting}

	This design consists of a decrementing counter that toggles a one-bit signal whenever it reaches 0. The one-bit signal triggers a LED which will blink. The blinking period is controlled by another signal. Following syntax elements can be noted:
	\begin{itemize}
		\item The component is created by inheriting from the \verb|Module| class. This is equivalent to a VHDL \verb|entity| or a Verilog \verb|module|.
		\item The basic element of any design is \verb|Signal|, which is similar to a VHDL \verb|signal| or a Verilog \verb|wire/reg|. Basic Python operations are redefined on signals, which means that expressions can be formed with a light syntax.
		\item The \verb|max| parameter of the \verb|Signal| constructor can be used to automatically determine width of multi-bit signals.
		\item The \verb|eq| method returns an assignment to a signal.
		\item As mentioned before, the design is split into combinatorial statements using the \verb|comb| attribute, and synchronous statements using the  \verb|sync| attribute.
		\item The \verb|If| object is used to represent conditional statements.
		\item In more complex designs, several modules can be integrated in one module using a \verb|submodules| property.
	\end{itemize}
	All signals in Migen are created using the same Signal object, and when
	they need to be accessed from outside, Migen converts them into Verilog
	I/O ports and automatically determines their direction.

	\subsection{MiSoC}
	LiteX originated from MiSoC\cite{Misoc} and reuses the key concepts and elements.

	While Migen offers the generation of digital logic with Python, MiSoC provides the SoC interconnect infrastructure and cores:
	\begin{itemize}
		\item Multiple CPU support: LatticeMico32, mor1kx, VexRiscv.
		\item Memory controller supports SDR, DDR, LPDDR, DDR2 and DDR3.
		\item HDMI video in/out cores (Spartan6).
		\item Provided peripherals: UART, GPIO, timer, GPIO, NOR flash/SPI flash controller, Ethernet MAC, and more.
		\item High performance and low resource usage.
		\item Portable and easy to customize thanks to Python- and Migen-based architecture.
		\item Design new peripherals using Migen and benefit from automatic CSR maps and logic, simplified DMAs, etc.
		\item Possibility to encapsulate legacy Verilog/VHDL code.
	\end{itemize}
	~\\
	MiSoC is used successfully by M-Labs to build all the gateware for ARTIQ\cite{Artiq} in a portable, flexible and easily maintainable way.

	\section{LiteX SoC builder, library and utilities}

	\begin{table}[!ht]
		\centering
		\caption{IP Components available in LiteX}
		\begin{tabular}{|l|l|}
			\hline
			IP name & note \\ \hline\hline

			LiteDRAM     &  DRAM core, fully pipelined, SDRAM to DDR3\\ \hline
			LiteEth      &  Ethernet core, PHYs, MAC, UDP/IP, up to 1Gbps\\\hline
			LitePCIe     &  DMA and MMAP PCIe core up to Gen2 X4 \\ \hline  % modified After Florent mail
			LiteSATA     &  SATA 1/2/3 core, DMA, RAID, Mirroring, up to 6Gbps\\ \hline
			LiteUSB      &  USB2.0/3.0 Slave FIFO core + DMA \\ \hline
			LiteSDCard   &  SD card core, DMA, up to UHS-1 (55MB/s R/W)\\ \hline
			LiteICLink   &  Comm core, IOserdes, Transceivers, up to 10gbps\\ \hline
			LiteJESD204B &  JESD204B core, TX, Subclass1, up to 10Gbps\\ \hline
			LiteVideo    &  Video core, HDMI RX/TX, Framebuffer, up to 1080p60\\ \hline
			LiteScope    &  Logic Analyzer core, access via various bus protocols\\ \hline

		\end{tabular}
		\label{table:litex1}
	\end{table}

	\begin{table}[!ht]
		\centering
		\caption{Softcores instantiable via MiSoC and LiteX.}

		\begin{tabular}{|l|l|}
			\hline
			Softcore & note \\ \hline\hline
			LM32*      &     Lattice Mico 32 bits\\ \hline
			Mor1kx*    &     OpenRISC 1000 compliant core\\ \hline
			PicoRV32  &    RISC-V core \cite{Picorv32}  \\ \hline
			VexRiscv*  &     RISC-V core \cite{VexRiscv}\\ \hline
		\end{tabular}
		~\\
		*: inherited from MiSoC.
		\label{table:cores}
	\end{table}

	Since 2015, LiteX has been evolving as a fork of MiSoC to provide more coherence for Enjoy-Digital's commercial projects
	and to ease collaboration with other open-source communities:

	\begin{itemize}
		\item More experimental features.
		\item High speed LiteXSim SoC simulator (based on Verilator).
		\item Wider collection of cores (PCIe, SATA, Ethernet, DRAM, HDMI, SDCard, USB, etc...).
		\item AXI4 support (MMAP/Stream/Lite).
		\item Debugging tools to control/analyze a SoC from Serial/Ethernet/PCIe.
		%\item etc...
	\end{itemize}

	An overview of LiteX is depicted on Figure \ref{litex}.

	LiteX library is made of a dozen of peripheral components, as well as softcores, summarized in Table \ref{table:litex1}.
	To the best of our knowledge, this gathering of SoC elements makes LiteX unique.
	LiteX components have been used in several open-source projects like HDMI2USB\cite{HDMI2USB}, Fupy\cite{Fupy}, NeTV2 \cite{NeTV2},
	Axiom SDI module \cite{Apertus}, PCIe Screamer \cite{PCIeScreamer} but also in various commercial applications.
	Softcores instantiable via LiteX are detailed in Table \ref{table:cores}.
	Among them, we find Lattice Mico 32 softcore (lm32), a 32-bits RISC architecture softcore, available for free with an open IP core licensing agreement.

	Beyond peripherals and softcores, LiteX also provides \verb|platform| descriptions, which extend the build library distributed with Migen.
	The import of a generic Xilinx is done using a single line of code. The specifics of this platform are then given following an object-oriented process.
	Next, the user can choose the naming of specific input-output pins (i/o) in an array and can finally instantiate a Python object.
	This platform object keeps track of the i/o mapping, the name of the FPGA target, as well as the actual toolchain used for bitstream generation.
	The following example describes the FPGA target present on a Digilent Nexys4DDR board.

	\begin{lstlisting}[frame=single,caption={Nexys4DDR platform description in Litex/Migen},captionpos=b]
from migen.build.xilinx import XilinxPlatform

_io = [
  ("user_led",  0, Pins("H17"), # etc),
  ("user_led",  1, Pins("K15"), # etc,
  # ... skipped for brevity
  ("serial", 0,
    Subsignal("tx", Pins("D4")),
    Subsignal("rx", Pins("C4")),
    IOStandard("LVCMOS33"),
  ),
  # ....skipped
]

class Nexys4DDR(XilinxPlatform):
  default_clk_name = "clk100"
  default_clk_period = 10.0

  def __init__(self):
    XilinxPlatform.__init__(self,
      "xc7a100t-CSG324-1", _io,
      toolchain="vivado")

platform = Nexys4DDR()
	\end{lstlisting}

	Concerning the application description itself, the underlying methodology remains similar.

	Once instantiated in Python, the user can request board pins easily :
	\begin{lstlisting}[frame=single]
user_leds = Cat(*
  [platform.request("user_led", i)
  for i in range(16)])
	\end{lstlisting}

	\section{Experiments with LiteX}
	\subsection{SoC synthesis using LiteX + Migen + Vivado}
	In this section, we describe a SoC built with LiteX and Vivado (for synthesis, place and route).

	\begin{figure*}[htb]
		\centering
		\includegraphics[scale=0.3]{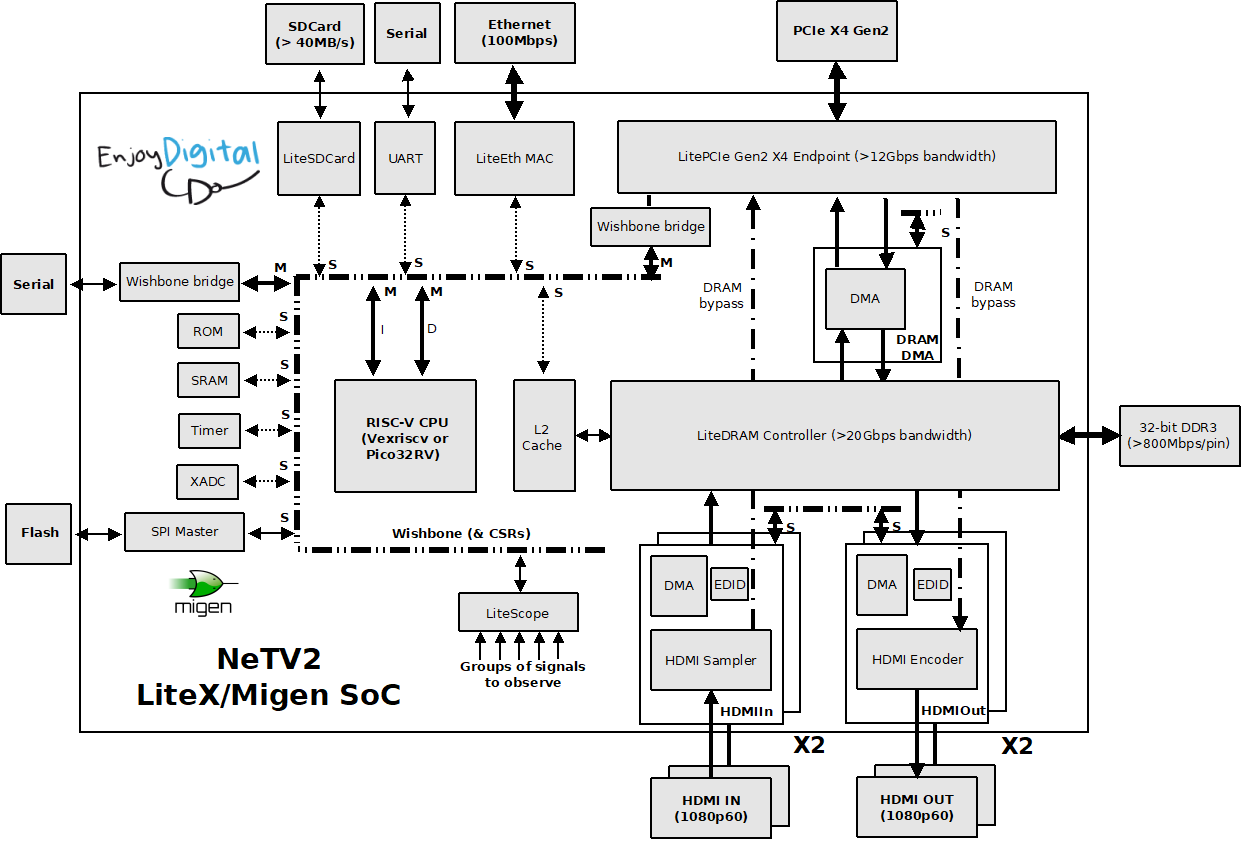}
		\caption{NeTV2 Libre SoC architecture}
		\label{netv2}
	\end{figure*}

	\begin{table}[!ht]
		\centering
		\caption{LiteX library reuse on NeTV2 Libre SoC}
		\begin{tabular}{|l|l|}
			\hline
			IP name & note \\ \hline\hline

			LiteVideo    &  Used to capture/play HDMI videos up to 1080p60\\
			&                                               \\ \hline
			LiteDRAM     &  Used to store/read video data to/from a 32-bits\\
			&  DDR3 (more than 20Gbps bandwidth)              \\\hline
			LitePCIe     &  Used to read/write data from/to the host at up \\
			&  to PCIe gen2 X4 (more than 12Gbps)             \\\hline
			LiteEth      &  Used to provide a 100Mbps control interface    \\
			&                                                 \\\hline
			LiteSDCard   &  Used to store CPU code and/or configuration data\\
			&                                                 \\\hline
		\end{tabular}
		\label{table:litex}
	\end{table}

	The NeTV2 Libre SoC is a variant of the official NeTV2 SoC (also using LiteX) and makes use of the
	NeTV2 hardware to create a PCIe capture/playback HDMI SoC with high debug capabilities. As can be seen on Figure \ref{netv2}, the SoC reuses LiteX capability to create complex SoCs and interconnect cores together,
	and also heavily reuses the LiteX library of open-source IPs (listed in Table \ref{table:litex}).
	The description of the top module in LiteX is too long to be given in this paper
	(but still under 1000 lines of code with the reuse of open-source IPs), but the following example (Listing \ref{lst:ddr3hdmi})
	describes how the DDR3 controller and HDMI out core are instantiated in the design.
	The synthesis of such a system using Xilinx Vivado took 10 minutes.

	\begin{lstlisting}[frame=single,label={lst:ddr3hdmi},caption={DDR3 and HDMI core instanciation using Litex/Migen},captionpos=b]
from litedram.modules import MT41J128M16
from litedram.phy import a7ddrphy
from litevideo.output import VideoOut

[...]

# sdram
self.submodules.ddrphy = a7ddrphy.A7DDRPHY(
  platform.request("ddram"), sys_clk_freq)
sdram_module = MT41J128M16(sys_clk_freq, "1:4")
self.register_sdram(self.ddrphy,
  sdram_module.geom_settings,
  sdram_module.timing_settings)

# hdmi out
hdmi_out_dram_port = self.sdram.crossbar.get_port(
  mode="read",
  dw=16,
  cd="hdmi_out0_pix",
  reverse=True)

self.submodules.hdmi_out0 = VideoOut(
  platform.device,
  platform.request("hdmi_out", 0),
  hdmi_out0_dram_port,
  "ycbcr422",
  fifo_depth=4096)
\end{lstlisting}

	\begin{itemize}
		\item The Artix7 DDRPHY is instantiated with \verb|ddram| pads and \verb|sys_clk| frequency passed to it.
		\item The DDRAM controller is created by registering the DDRAM PHY and DDRAM Module to the SoC.
		\item A DDRAM port is created on the controller for the HDMI output in \verb|read mode|, with a \verb|dw| of 16-bits (the logic to convert to
		native DDRAM width is automatically inserted) and with a specific \verb|clock domain| (the CDC logic is automatically inserted).
		\item The video out core is created by passing the HDMI pads, DDRAM port and configuration parameters to the VideoOut core.
	\end{itemize}
	%~\\
	This demonstrates that instantiating complex cores can be done in an efficient way and that many of the traditional RTL design tasks
	are handled automatically. For instance, the DDRPHY controller adapts itself to the provided DDRAM pads and module.
	Bus width conversion and CDC (clock domain crossing) can also be handled automatically.
	%~\\
	LiteX generates the RTL design and placement/timing constraints files for this SoC. % in a few seconds.
	Vivado then handles the synthesis, place and route.% in a few minutes.

	\subsection{SoC synthesis using LiteX + Migen + Trellis + Yosys + Nextpnr}
	In this section, we show how LiteX can be used as an entry to create a SoC for a Lattice ECP5 FPGA
	board with a fully open-source toolchain. Three main projects are used here: Trellis project \cite{Trellis}, which aims at documenting ECP5 FPGAs bitstream, Yosys \cite{Yosys} for Verilog synthesis, and Nextpnr, a vendor neutral,
	timing driven, FPGA place and route tool.
	%~\\
	The following code (Listing \ref{lst:soc_lm32}) creates a SoC with a LM32 CPU, ROM, UART, Timer, SDRAM controller that executes
	its BIOS on startup and can then run code from SDRAM when loaded over serial link. The code makes an explicit
	call to the toolchain (named ``trellis''), just the same way as for ``vivado''.
	The full synthesis (bitstream included), starting from LiteX, took 14 minutes, which is the same order of magnitude as for the commercial toolchain experiment.

	\begin{lstlisting}[frame=single,label={lst:soc_lm32},caption={LM32-based SoC Using Litex/Migen coupled with a full open-source EDA stack.},captionpos=b]
from migen import *
from litex.boards.platforms import versa_ecp5
from litex.soc.cores.clock import *
from litex.soc.integration.soc_sdram import *
from litex.soc.integration.builder import *
from litedram.modules import AS4C32M16
from litedram.phy import GENSDRPHY


class _CRG(Module):
  def __init__(self, platform):
    self.clock_domains.cd_sys = ClockDomain()
    self.clock_domains.cd_sys_ps = ClockDomain()
    # clk / rst
    clk100 = platform.request("clk100")
    rst_n = platform.request("rst_n")
    platform.add_period_constraint(clk100, 10.0)
    # pll
    self.submodules.pll = pll = ECP5PLL()
    self.comb += pll.reset.eq(~rst_n)
    pll.register_clkin(clk100, 100e6)
    pll.create_clkout(self.cd_sys, 50e6)
    pll.create_clkout(self.cd_sys_ps, 50e6,
    phase=90)
    self.comb += self.cd_sys.rst.eq(~rst_n)
    # sdram clock
    sdram_clock = platform.request("sdram_clock")
    self.comb += sdram_clock.eq(sdram_ps_clk)

class BaseSoC(SoCSDRAM):
  def __init__(self, **kwargs):
    platform = versa_ecp5.Platform(
      toolchain="trellis")
    sys_clk_freq = int(50e6)
    SoCSDRAM.__init__(self, platform,
      cpu_type="lm32", l2_size=32,
      clk_freq=sys_clk_freq,
      integrated_rom_size=0x8000)

    self.submodules.crg = _CRG(platform)

    self.submodules.sdrphy = GENSDRPHY(
      platform.request("sdram"))
    sdram_module = AS4C32M16(sys_clk_freq, "1:1")
    self.register_sdram(self.sdrphy,
      sdram_module.geom_settings,
      sdram_module.timing_settings)

soc = BaseSoC()
builder = Builder(soc)
builder.build()
	\end{lstlisting}

	\begin{figure}[htb]
		\centering
		\includegraphics[scale=0.1]{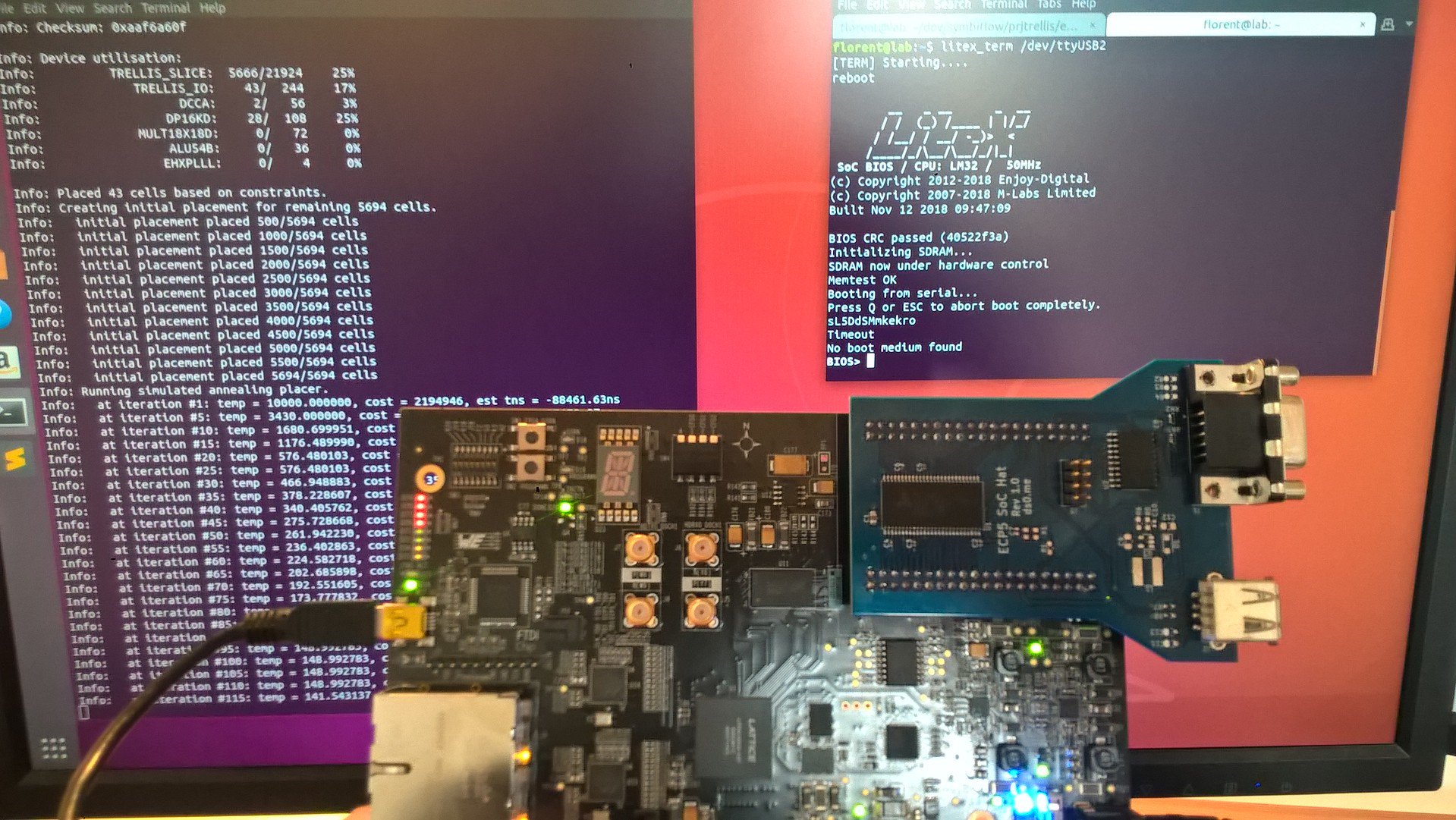}
		\caption{Versa ECP5 + SDRAM SoC Hat with LiteX / LiteDRAM}
		\label{versa_ecp5}
	\end{figure}
	%~\\
	This example proves that LiteX entry can be used with a fully open-source toolchain.
	Lattice iCE40 and ECP5 FPGAs are already supported. Furthermore, these tools have been designed to be portable to
	others FPGA families. Communities are actually documenting other FPGA families, so we could expect
	having fully open-source toolchains for Xilinx and Altera FPGAs in the next couple of years.
	% Open-source toolchains offers lots of new possibilities:
	% \begin{itemize}
	%   \item Build times of a few seconds on simple designs: less than 4 seconds measured from
	%   start to a programmed FPGA with the LiteX / Migen / Trellis / Yosys / Nextprn flow...
	%   \item Look at the code, fix yourself the issue within hours instead of months or years with
	%   vendor tools when bug is reported (when even fixed...) or generally having to find an acceptable workaround.
	%   \item Make FPGA easily accessible and to everyone.
	% \end{itemize}

	\section{Conclusion and perspectives}
	In this paper, we presented LiteX library together with its key DSL: Migen.
	As an open-source library, made of several processors and peripherals, LiteX
	has proven efficient and robust enough for the design of complex system-on-chip.
	LiteX is especially well coupled with FPGAs: Python ease of use seems to add a new argument
	to a wider adoption of FPGAs. Open-source toolchains will probably bring many new innovations
	in the future and coupled with Migen, MiSoC/LiteX or others DSLs, have already enabled new ways of
	thinking and designing hardware.

	\bibliographystyle{ieeetr}
	\bibliography{dsl_eda}
	%\nocite{label}
	\nocite{*}

\end{document}